\documentclass[
prl,
superscirptaddress,
showpacs,
amsmath,
amssymb,
a4paper,
twocolumn
]{revtex4}

\usepackage[hmarginratio=1:1,top=3cm,height=24.7cm,width=18cm,a4paper]{geometry}

\usepackage{graphicx}
\usepackage{dcolumn}
\usepackage{bm}
\begin{document}

\preprint{APS/123-QED}

\title{Dynamics of coherent and incoherent emission from an artificial atom in a 1D space}

\author{A.~A.~Abdumalikov,~Jr.$^1$}
\email{abdumalikov@phys.ethz.ch}
\altaffiliation[On leave from ]{Physical-Technical Institute,
Tashkent 100012, Uzbekistan}
\altaffiliation[Present address ]{Department of Physics, ETH Zürich, CH-8093, Zürich, Switzerland}
\author{O.~V.~Astafiev$^{1,2}$}
\email{astf@zb.jp.nec.com}
%\author{Yu.~A.~Pashkin$^{1,2}$}
%\altaffiliation[On leave from]{Lebedev Physical Institute, Moscow 119991, Russia}
\author{Y.~Nakamura$^{1,2}$}
\author{J.~S.~Tsai$^{1,2}$}
\affiliation{
$^{1}$RIKEN Advanced Science Institute, Wako, Saitama 351-0198, Japan\\
$^{2}$NEC Green Innovation Research Laboratories, Tsukuba, Ibaraki 305-8501, Japan
}

\date{\today}

\begin{abstract}
We study dynamics of an artificial two-level atom in an open 1D space by measuring evolution of its coherent and incoherent emission. States of the atom -- a superconducting flux qubit coupled to a transmission line -- are fully controlled by resonant excitation microwave pulses. The coherent emission -- a direct measure of superposition in the atom -- exhibits decaying oscillations shifted by $\pi/2$ from oscillations of the incoherent emission, which, in turn, is proportional to the atomic population. The emission dynamics provides information about states and properties of the atom. By measuring the coherent dynamics, we derive two-time correlation function of fluctuations and, using quantum regression formula, reconstruct the incoherent spectrum of the resonance fluorescence triplet, which is in a good agreement with the directly measured one.
\end{abstract}

\maketitle

Superconducting circuits with small Josephson junctions are now commonly used to demonstrate a variety of quantum phenomena. The circuits were shown to possess discrete energy levels, controlled quantum coherent evolution \cite{Nakamura1999,Martinis2002,Vion2002,Chiorescu2003} and coherent interaction with quantized electromagnetic modes of resonators \cite{Wallraff2004,Abdumalikov2009}. In a series of experiments many fundamental effects from quantum optics
have been demonstrated \cite{Schuster2007,Fragner2008,Astafiev2007,Neeley2009,Astafiev2010,Astafiev2010a,Abdumalikov2010}. Strong interaction of the circuits with the electromagnetic fields allows also to demonstrate new phenomena, e.g. controllable manipulations with single photons and photon fields \cite{Houck2007,Hofheinz2009}.

Recently we have realized an analog of a natural atom in the open space -- an artificial atom in the 1D space \cite{Astafiev2010}. The atom is strongly coupled to a transmission line via dipole interaction. Here we study dynamics of coherent and incoherent dipole emission from such an atom. Manipulating the atomic states by microwave pulses, one can perform the quantum state and process tomography. Differently from the qubit tomography realized by projective measurement to its states \cite{Tomography}, we perform essentially optical measurements of the dipole emission from the non-isolated atom. The excited atom is a dynamical system, which continuously emits radiation to the line, similarly to natural atoms in the open space. However differently from the optical measurements of the natural atoms, the collection efficiency of the emission in the 1D transmission line is very high due to its strong coupling to the line \cite{Astafiev2010}. We suggest a convenient experimental procedure for extracting the two-time correlation function of fluctuations, using the time-dependences of the coherent dipole emission. The incoherent spectrum of resonance fluorescence is then calculated, which shows a good agreement with the experimentally measured one.

As an artificial atom, we use a superconducting flux qubit \cite{Mooij1999,Chiorescu2003} inductively coupled to a coplanar transmission line \cite{Astafiev2010}. The two lowest eigenstates of the atom (qubit states) are formed by the states associated with clockwise and counterclockwise circulating currents with the persistent current $I_p=213\ $nA. Their energies are controlled by the external magnetic flux threading through the loop $\Phi=\Phi_0/2+\delta\Phi$, where $\delta\Phi$ is the deviation from the half-flux quantum $\Phi_0/2$. We tune the flux bias to the optimal point $\delta\Phi=0$, where the atomic transition frequency is $\omega_a/2\pi = 9.888\ $GHz. The loop shares a part with the transmission line, which results in a mutual inductance $M=13.6\ $pH mainly due to the kinetic inductance of the shared segment.

The two-level atom driven by the resonant microwave, with the current field $I_0 \cos{(\omega_a t - \varphi)}$, is described by the Hamitonian $H = -\hbar\Omega (\sigma^+e^{i\varphi} + \sigma^-e^{-i\varphi})/2$ in the rotating wave approximation. Here $\sigma^\pm = (\sigma_x \pm i\sigma_y)/2$, and $\sigma_x$, $\sigma_y$, $\sigma_z$ are the Pauli matrices, $\Omega=\phi_pI_0/(2\hbar)$ is the Rabi frequency, and $\phi_p = M I_p$ is the atomic dipole transition matrix element. The dynamics of such a system is described by a spin-1/2 in the magnetic field \cite{Slichter} and is governed by the optical Bloch equations
\begin{equation}\label{Bloch}
\frac{d\vec{\sigma}}{dt} = \mathbf{B}\vec{\sigma} + \vec{b},
\end{equation}
where
\begin{equation}\label{Bloch_not}
\mathbf{B}=\left(
\begin{array}{ccc}
-\Gamma_2  & 0 &  -\Omega \sin\varphi\\
0  & -\Gamma_2 &  -\Omega\cos\varphi\\
\Omega\sin\varphi  & \Omega\cos\varphi  & -\Gamma_1
\end{array}
\right),
\end{equation}
$\vec{b} = \{0,0,-\Gamma_1\}$, and $\vec{\sigma} = \{\langle\sigma_x\rangle, \langle\sigma_y\rangle, \langle\sigma_z\rangle\}$ is a vector of the expectation values of the Pauli matrices. This vector represents the atomic state, accounting incoherent processes of relaxation with the rate $\Gamma_{1}$ (decay of $z$-component) and dephasing with the rate $\Gamma_2=\gamma+\Gamma_1/2$ (decay in $xy$-plane), where $\gamma$ is the pure dephasing rate. In the Rabi rotation, the phase $\varphi$ controls the axis; e.g., $\varphi = 0$ and $\varphi = \pi/2$ cause the spin rotations around $x$ and $y$ axes, respectively. Combining pulses with different $\varphi$, one can controllably rotate the spin, while $|\vec{\sigma}| < 1$ subjected to the incoherent processes.

The atom generates two coherent waves propagating forward and backward with the current field \cite{Astafiev2010}
\begin{equation}\label{current}
I^{\mp}(x, t)=\frac{\hbar\Gamma_1}{\phi_p}\langle i\sigma^\pm\rangle e^{ik|x|-i\omega t},
\end{equation}
found from the expectation value of the current operator $\widehat{I}^{\mp} = (\hbar\Gamma_1/\phi_p)i\sigma^\pm$ of the atom situated at $x = 0$. Equation (\ref{current}) presents the dipole emission proportional to $\langle \sigma^\pm\rangle$, that is, to the projection of the pseudo-spin on the $xy$-plane. Therefore, the emission amplitude is a direct measure of superposition in the atom and reaches maximum ($|\langle\sigma^\pm\rangle| = 1/2$) in a maximally superposed state.
The real and imaginary parts of the emitted wave (0- and 90-degree phase in respect to the driving wave) give $\langle\sigma_y\rangle$ and $\langle\sigma_x\rangle$, respectively.
In the case of the atom driven by the microwave with $\varphi = 0$ from its ground state $\{0,0,-1\}$, the emitted wave is purely real because $\langle\sigma_x\rangle = 0$.

\begin{figure}[tbp]
\includegraphics{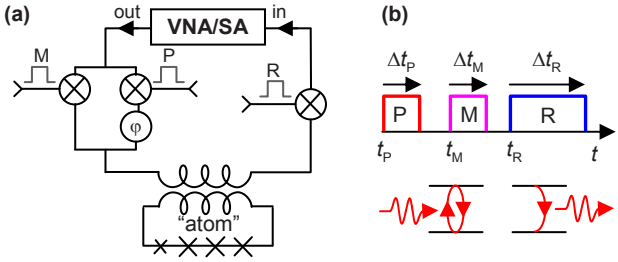}
\caption{\label{fig1} (a) Measurement circuit diagram. (b) General pulse sequence (upper panel) with schematic atomic dynamics (lower panel): The pulses ${\rm P}$ and ${\rm M}$ are used to prepare and manipulate the atomic states, and the emission from the atom is detected during the readout pulse ${\rm R}$.
}
\end{figure}

In Fig.~\ref{fig1}(a), a schematic diagram of the measurement setup is presented. We prepare and manipulate the atom by one or two sequential microwave pulses (denoted by ${\rm P}$ and ${\rm M}$) of lengths $\Delta t_{\rm P}$ and $\Delta t_{\rm M}$ applied at times $t_{\rm P}$ and $t_{\rm M}$
[Fig.~\ref{fig1}(b)]. A continuous microwave is split into two channels and chopped independently by the rectangular pulses ${\rm P}$ and ${\rm M}$ with the choppers consisting of high-frequency mixers. The microwave in the ${\rm P}$ channel acquire a phase shift $\varphi$ (=0 unless it is specified) produced by a computer-controlled mechanical phase shifter. The two channels are combined into one, and the signal is delivered to the sample in a dilution refrigerator (at $T \approx$~50~mK) through a coaxial cable.

The output signal is amplified by a cryogenic amplifier and a room-temperature amplifier and then chopped by the readout pulse ${\rm R}$ of a length $\Delta t_{\rm R}=50$ ns ($\Delta t_{\rm R} \gg \Gamma_2^{-1}$) at time $t_{\rm R}$ after the manipulation is completed. The signal is detected either by a vector network analyzer (VNA) in the homodyne measurement of the coherent emission or by a scalar spectrum analyzer (SA) in the power measurement. As it follows from Eqs.~(\ref{Bloch}) and (\ref{current}), during the readout pulse ${\rm R}$ the current amplitude of the coherent emission is $I^+(t) = (\hbar\Gamma_1/\phi_p)\langle\sigma^-(t_{\rm R})\rangle e^{-\Gamma_2 (t-t_{\rm R})}$ and the forward (backward) emitted power is $P(t) = (\hbar\omega_a\Gamma_1/4)(1+\langle\sigma_z(t_{\rm R})\rangle) e^{-\Gamma_1 (t-t_{\rm R})}$. The signal is integrated over $\Delta t_{\rm R}$ and the pulse sequence is repeated with a period of $T_r = 250$ ns, resulting in the averaged detected signals by VNA and SA $I^+ = \hbar\Gamma_1/(\phi_p\Gamma_2 T_r)\langle\sigma^-(t_{\rm R})\rangle$ and $P = \hbar\omega_a/(4T_r)(1+\langle\sigma_z(t_{\rm R})\rangle)$, respectively. We will omit $t_{\rm R}$ in further notations.

\begin{figure}[tbp]
\includegraphics[scale=1.0]{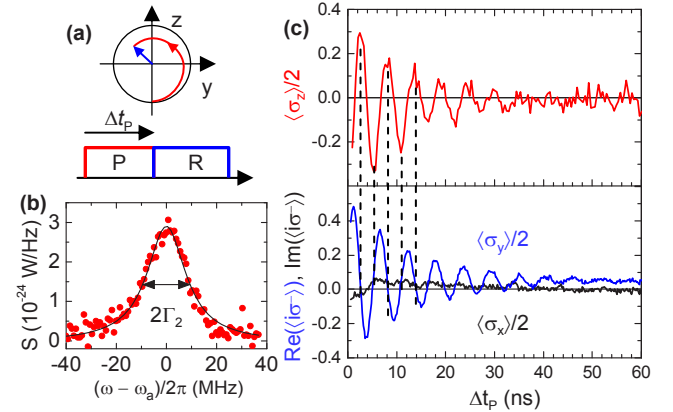}
\caption{\label{fig2} (a) Pulse sequence for the preparation and readout the pseudo-spin and the corresponding rotation in $yz$-plane of the Bloch sphere. (b) Power spectrum of the free-induction decay emission peak with a long excitation pulse $\Delta t_{\rm P}$ = 100~ns. (c) Evolution of the atomic population measured by the emitted power (upper panel), and the normalized dipole moment $\langle\sigma^-\rangle$ derived from the coherent dipole emission (lower panel). The dashed lines mark the points where the emission is completely incoherent ($\langle\sigma^-\rangle = 0$), corresponding to the extrema of the emitted power ($\propto 1+\langle \sigma_z \rangle$).}
\end{figure}

Coherent and incoherent dynamics of the system is studied by measuring the emission from the atom. A single microwave pulse ${\rm P}$ of varied length $\Delta t_{\rm P}$ is applied to the atom with the readout pulse ${\rm R}$ following right after ${\rm P}$ as shown in Fig.~\ref{fig2}(a). The SA spectrum of the incoherent emission from the atom excited by a long pulse with $\Delta t_{\rm R}$~=~100~ns (to saturate its population to 50\%) is shown in Fig.~\ref{fig2}(b) by the red dots. By fitting it with a Lorentzian (black curve), we obtain the dephasing rate $\Gamma_2/2\pi$ = 9.4~MHz. The peak power contains information about the atomic population. The upper panel in Fig.~\ref{fig2}(c) shows $\langle\sigma_z\rangle$, measured by the emitted power in 5~MHz bandwidth at $\omega_a$. On the other hand, the lower panel shows the dynamics of $\langle\sigma^-\rangle$ derived from the coherent dipole emission measured by VNA. The behavior is consistent with the rotation of the spin in $yz$-plane: ${\rm Re}(\langle i\sigma^-\rangle) = \langle\sigma_y\rangle/2$ (blue curve) exhibits decaying oscillations, while the imaginary part ${\rm Im}(\langle i\sigma^-\rangle) = \langle\sigma_x\rangle/2$ (black curve) remains nearly zero.
The total emission becomes completely incoherent in the extrema marked by dashed lines, because at these points $\langle\sigma^-\rangle = 0$. The curves of Fig.~\ref{fig2}(c) have also been used for calibration.

\begin{figure}[tbp]
\includegraphics[scale=1]{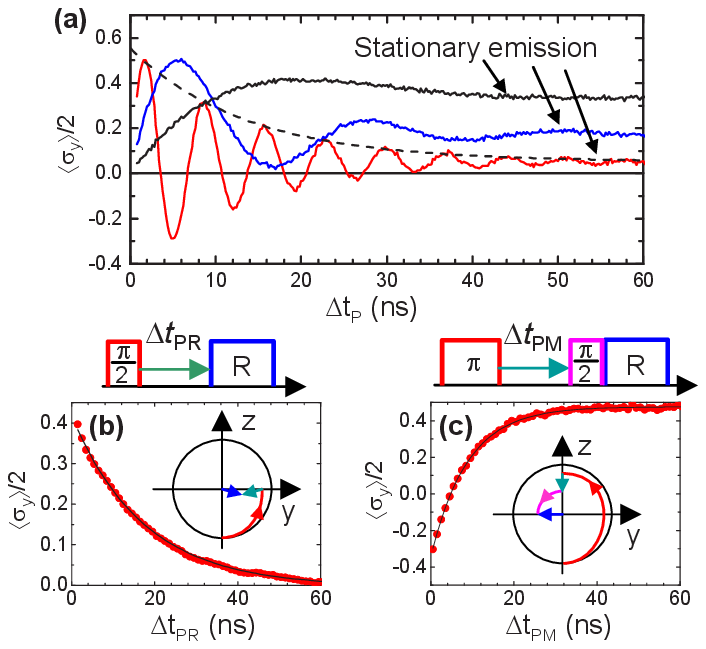}
\caption{\label{fig3} (a) Evolution of $\langle\sigma_y\rangle/2$ at different driving amplitudes $\Omega/2\pi$ = 140~MHz (red curve), 44~MHz (blue curve) and 14~MHz (black curve). The stationary value of $\langle\sigma^-\rangle$ strongly depends on the driving power. The dashed black curve depicts decay of the fast oscillations with rate $(\Gamma_1/2+\Gamma_2/2)/2\pi$ = 13.5~MHz.  (b) Direct measurements of dephasing.
The red dots show $\langle\sigma_y\rangle$ as a function of the delay $\Delta t_{\rm PR}$ between the manipulation $\pi/2$ and readout pulses. The black curve is an exponential-decay fit with $\Gamma_2/2\pi = 9.1$~MHz (c) Measurements of relaxation.
The measured decay of $\langle\sigma_y\rangle$ (projected $\langle\sigma_z\rangle$) is presented by the red dots. It changes the sign when $\langle\sigma_z\rangle$ becomes negative. The black curve is an exponential fit with $\Gamma_1/2\pi = 18.3$~MHz.
}
\end{figure}

The oscillations decay, while the emission saturates to a finite stationary level defined by $\langle i\sigma^-\rangle = (\Gamma_1\Omega/2)/(\Gamma_1\Gamma_2+\Omega^2)$ ($\equiv\langle\sigma_y\rangle/2$). The stationary emission at different driving amplitudes is exemplified in Fig.~\ref{fig3}(a), where the evolutions of $\langle\sigma_y\rangle/2$ are shown for $\Omega/2\pi =$ 140~MHz (red curve), 44~MHz (blue), and 14~MHz (black). Here, the saturation (stationary) level increases with decreasing $\Omega$. As it follows from Eq.~(\ref{Bloch}), the oscillations decay with the rate $\Gamma_1/2+\Gamma_2/2$, because the rotating pseudo-spin is subjected equally to relaxation in $z$-axis as well as dephasing in $y$-axis. The dashed black curve tracing the oscillation maxima is the exponential decay with $(\Gamma_1/2+\Gamma_2/2)/2\pi \approx 13.5$~MHz.

Having full control of our system, $\Gamma_1$ and $\Gamma_2$ can be measured separately. Figures.~\ref{fig3}(b) and (c) present relaxation processes with decay rates $\Gamma_2$ and $\Gamma_1$. To determine $\Gamma_2$, we apply a $\pi/2$-${\rm P}$-pulse [the lengths of $\pi/2$- and $\pi$-pulses are defined from the first maximum and first zero on the blue curve of Fig.~\ref{fig2}(c)] which brings the pseudo-spin to $y$-axis, and then measure the emission after a delay $\Delta t_{\rm PR}$. The decay of $\langle\sigma_y\rangle$ as a function of $\Delta t_{\rm PR}$ is shown in Fig.~\ref{fig3}(b). To measure the energy relaxation, we apply $\pi$-${\rm P}$-pulse for preparing the pseudo-spin up (along $z$-axis), and after a delay $\Delta t_{\rm PM}$, during which the population $(1+\langle\sigma_z\rangle)/2$ decays, we apply $\pi/2$-$\rm M$-pulse rotating the spin from $z$- to $y$-direction, as shown in Fig.~\ref{fig3}(c). By fitting the experimentally measured emissions with exponential decays, we obtain $\Gamma_1/2\pi = 18.3$~MHz and $\Gamma_2/2\pi = 9.1$~MHz ($\Gamma_2\approx\Gamma_1/2$), which indicates that pure dephasing is negligible in the system. These numbers are consistent with the ones separately determined above.

With the demonstrated state manipulations, one could perform in principle the quantum state and process tomography, obtaining full information about the time evolution of the atom.
Here, instead, we demonstrate derivation of incoherent properties of our system by measuring only the coherent emission. Namely, we reconstruct the two-time correlation function of fluctuations, derive incoherent spectrum of the resonance fluorescence (Mollow) triplet and compare it with the spectrum directly measured by SA. To derive the correlation function the full set of tomography measurements is not needed. We propose a convenient procedure for the correlation function derivation from experimental curves of the coherent dipole emission.

The incoherent emission power spectral density in either direction is given by $S(\omega) = (1/2\pi) Z\int_{-\infty}^{\infty}\langle\Delta\widehat{ I}^{-}(0)\Delta\widehat{I}^{+}(t)\rangle_{\rm ss} e^{i\omega t} dt $, where $\Delta \widehat{o} = \widehat{o} - \langle \widehat{o}\rangle_{\rm ss}$ denotes deviation (fluctuations) of the operator $\widehat{o}$ from its steady state $\langle \widehat{o}\rangle_{\rm ss} = \lim_{t\to\infty}\langle \widehat{o}(t)\rangle$ \cite{Carmichael} and $Z$ ($=50$~$\Omega$) is the characteristic impedance of the transmission line. Substituting $\widehat{I}^{\mp}$ from Eq.~(\ref{current}) and using the atomic relaxation rate $\Gamma_1 = \hbar\omega_a\phi_p^2/(\hbar^2 Z)$ into the two directions along the line, we arrive at
\begin{equation}\label{spectr}
S(\omega)=\frac{\hbar\omega_a\Gamma_1}{2\pi}\int\limits_{-\infty}^{\infty} \langle \Delta\sigma^+(0)\Delta\sigma^-(t)\rangle_{\rm ss} e^{i\omega t} dt.
\end{equation}

According to the quantum regression formula \cite{Carmichael} the steady state of the two-time correlation function $\langle\Delta\sigma^+(0)\Delta\sigma^-(t)\rangle_{\rm ss}$ can be found by solving the equation
\begin{equation}\label{cor}
\frac{d\vec{s}(t)}{dt}={\mathbf B}\vec{s}(t)
\end{equation}
for $\vec{s}(t) = \langle\Delta\sigma^+(0)\Delta\vec{\sigma}(t)\rangle_{\rm ss}$ (where $\vec{s} = \{s_x,s_y,s_z\}$) with the initial conditions $\vec{s}(0)$ obtained from the steady-state $\vec{\sigma}_{\rm ss}$ and then by calculating $\langle\Delta\sigma^+(0)\Delta\sigma^-(t)\rangle_{\rm ss} = [s_x(t) - i s_y(t)]/2$. For the strong drive ($\Omega\gg \Gamma_2$) where $\langle\sigma_{x,y,z}\rangle_{\rm ss} \rightarrow 0$, the initial conditions are simplified to
\begin{equation}\label{init}
\vec{s}(0) \approx
\frac{1}{2}
\left(
\begin{array}{c}
1\\
i\\
0
\end{array}
\right)
.
\end{equation}
The solution for $\vec{s}(t)$ can be found from $\Delta\vec{\sigma}(t) = \vec{\sigma}(t)-\vec{\sigma}_{\rm ss}$ by measuring the coherent dynamics of $\vec{\sigma}(t)$, governed by Eq.~(\ref{Bloch}) for two initial conditions $\vec{\sigma}'(0) = \{1,0,0\}$ and $\vec{\sigma}''(0) = \{0,1,0\}$ and subtracting $\vec{\sigma}_{\rm ss} \approx \vec{\sigma}(t)$ at $t \gg \Gamma_2$.

Here, however, we demonstrate an equivalent but slightly different approach, which is more practical experimentally. We measure evolution of $\vec{\sigma}(t)$
with two pairs of opposite initial conditions:
$\vec{\sigma}'_{(\pm)}(0) = \{\pm 1,0,0\}$ and
$\vec{\sigma}''_{(\pm)}(0) = \{0,\pm 1,0\}$ prepared by applying $\pi/2$-${\rm P}$-pulses with $\varphi = \pi \mp \pi/2$ and $\varphi = \pi/2 \mp \pi/2$, respectively.
Differences in the pairs give solutions of Eq.~(\ref{cor}):
$\vec{s}'(t) = [\vec{\sigma}'_{(+)}(t)-\vec{\sigma}'_{(-)}(t)]/2$ for $\vec{s}'(0) = \{1,0,0\}$ and
$\vec{s}''(t) = [\vec{\sigma}''_{(+)}(t)-\vec{\sigma}''_{(-)}(t)]/2$ for $\vec{s}''(0) = \{0,1,0\}$. The initial conditions of Eq.~(\ref{init}) can be rewritten as $\vec{s}(0) = [\vec{s}'(0) + i \vec{s}''(0)]/2$ and, therefore, the desired correlation function is $\langle\Delta\sigma^+(0)\Delta\sigma^-(t)\rangle_{\rm ss} = \{[s'_x(t)-i s'_y(t)]+i[s''_x(t)-i s''_y(t)]\}/4$. Such a procedure provides robustness against small errors in $\varphi$ and does not require separate measurements of $\vec{\sigma}_{\rm ss}$.

\begin{figure}[tbp]
\includegraphics[scale=1.0]{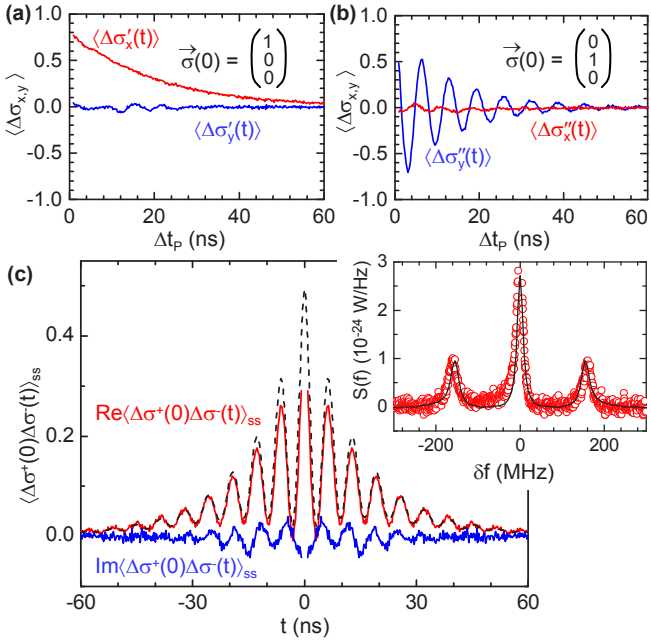}
\caption{\label{fig4} (a) and (b) Evolutions of $\langle\Delta\sigma_{x,y}(t)\rangle$ with initial conditions $\vec{\sigma}(0)=\{1,0,0\}$ and $\{0,1,0\}$, respectively. (c) Two-time correlation function of fluctuations $\langle\Delta\sigma^+(0)\Delta\sigma^-(t)\rangle_{\rm ss}$ calculated based on evolutions in (a) and (b). The red and blue are the real and imaginary parts, and the dashed black curve is a real part of the solution obtained from Eq.~(5) using {\bf{B}} with experimentally deduced parameters. In the experiment, the points at times shorter than 0.8~ns could not be measured. Inset: Spectrum [$S(f) = 2\pi S(\omega)$] of the resonance fluorescence triplet as a function of detuning $\delta f = (\omega-\omega_a)/2\pi$ derived from the two-time correlation function according to Eq.~(\ref{spectr}) (black curve) and measured directly in the frequency domain SA (red open circles).
}
\end{figure}

The evolutions of the $x, y$-components of $\Delta\vec{\sigma}'$ and $\Delta\vec{\sigma}''$
%found from subtraction of experimentally measured $\langle\sigma^-(t)\rangle$ for $\{\pm 1,0,0\}$ and $\{0,\pm 1,0\}$
for $\vec{\sigma}(0) = \{1,0,0\}$ and $\{0,1,0\}$ are shown in Figs.~\ref{fig4}(a) and (b). The real and imaginary parts of the correlation function $\langle\Delta\sigma^+(0)\Delta\sigma^-(t)\rangle_{\rm ss}$ derived from those plots are  shown in Fig.~\ref{fig4}(c) by the red and blue curves. The dashed black curve shows the correlation function alternatively calculated from Eq.~(\ref{cor}) using {\bf B} with $\Gamma_1$ and $\Gamma_2$. The correlation function of a superconducting quantum circuit has also been recently measured more directly in Ref.~\cite{Bozyigit2010}.
By taking Fourier transformation of the correlation function [see Eq.~(\ref{spectr})], we obtain the spectrum of resonance fluorescence, which coincides well with the experimentally measured one as shown in the inset of Fig.~\ref{fig4}(c).

In conclusion, we have studied time-evolution of coherent and incoherent dipole emission from an artificial two-level atom strongly coupled to an open transmission line. The correspondence between coherent and incoherent emission has been shown. We demonstrate derivation of the two-time correlation function and reconstruction of the spectrum of resonance fluorescence (Mollow) triplet from time-dependence of the coherent emission.

This work was supported by Funding Program for World-Leading Innovative R\&D on Science and Technology (FIRST), MEXT KAKENHI ``Quantum Cybernetics'' and JST Core Research for Evolutional Science and Technology projects.

%\bibliography{Tomography}

\end{document}